\documentclass[conference]{IEEEtran}
\IEEEoverridecommandlockouts
\usepackage{cite}
\usepackage{amsmath,amssymb,amsfonts}
\usepackage{graphicx}
\usepackage{algorithmic}
\usepackage[]{algorithm2e}
\usepackage{textcomp}
\usepackage{xcolor}
\usepackage{multirow}
\usepackage{paralist}
\def\BibTeX{{\rm B\kern-.05em{\sc i\kern-.025em b}\kern-.08em
    T\kern-.1667em\lower.7ex\hbox{E}\kern-.125emX}}
\begin{document}
\makeatletter
    \newcommand{\linebreakand}{%
      \end{@IEEEauthorhalign}
      \hfill\mbox{}\par
      \mbox{}\hfill\begin{@IEEEauthorhalign}
    }
\makeatother
\title{Analyzing the Productivity of GitHub Teams\\ based on Formation Phase Activity}

\author{\IEEEauthorblockN{Samaneh Saadat}
\IEEEauthorblockA{\textit{Department of Computer Science} \\
\textit{University of Central Florida}\\
Orlando, FL US \\
ssaadat@cs.ucf.edu}
\and
\IEEEauthorblockN{Olivia B. Newton}
\IEEEauthorblockA{\textit{School of Modeling, Simulation, and Training} \\
\textit{University of Central Florida}\\
Orlando, FL US \\
olivianewton@knights.ucf.edu}
\linebreakand
\IEEEauthorblockN{Gita Sukthankar}
\IEEEauthorblockA{\textit{Department of Computer Science} \\
\textit{University of Central Florida}\\
Orlando, FL US \\
gitars@eecs.ucf.edu}
\and
\IEEEauthorblockN{Stephen M. Fiore}
\IEEEauthorblockA{
\textit{Department of Philosophy} and \\
\textit{School of Modeling, Simulation, and Training} \\
\textit{University of Central Florida}\\
Orlando, FL US \\
sfiore@ist.ucf.edu}
}
\maketitle

\begin{abstract}
Our goal is to understand the characteristics of high-performing teams on GitHub.  
Towards this end, we collect data from software repositories and evaluate teams by examining differences in productivity.  Our study focuses on the team formation phase, the first six months after repository creation.
To better understand team activity, we clustered repositories based on the proportion of their work activities and discovered three work styles in teams: \textit{toilers}, \textit{communicators}, and \textit{collaborators}.
%%%%%%%%%%
% For our prediction model, we identify and extract a set of features based on activity during the first six months after repository creation.
% These features were used to predict the productivity of teams a year after repository creation.
% Our model is capable of predicting the performance of software development teams with an F1 score of 0.89 and can be used to forecast the future trajectory of software development processes based on team formation phase activity.
Based on our results, we contend that early activities in software development repositories on GitHub establish coordination processes that enable effective collaborations over time.
\end{abstract}


\section{Introduction}

GitHub\footnote{https://github.com/}, a social coding platform used by self-organized teams for open source software development (OSSD), is increasingly popular among researchers as a source for data mining of software repositories.
Recent work in this research domain includes the prediction of repository popularity based on historical data \cite{Borges:2016:popular}, contribution choices of developers \cite{Nielek2016:devchoice}, and file changes based on communication networks \cite{Wiese2015:changes}.
Many of these studies have collected data from mature, large software repositories to predict outcomes of interest.
To add to this body of work, in the present study we examine a variety of repositories from the early stages of creation to better understand the relationship between processes in team formation and subsequent performance.  The paper addresses the following research questions:
\begin{compactitem}
\item \textbf{RQ1:} how does team size and work centralization differ during the formation period?
\item \textbf{RQ2:} how is work style related to team performance? 
\item \textbf{RQ3:} which team activity features are closely related to  performance?
\end{compactitem}

In group research, team formation describes the foundational processes that support subsequent team interactions.
Our conceptualization of team formation is based on Kozlowski et al.'s normative model of team development \cite{Kozlowski:teamdev}.
This model defines four continuous and overlapping phases: team formation, task compilation, role compilation, and team compilation.
In the team formation phase, team members begin to learn about each other and establish norms and shared goals through information-seeking behaviors, and thus develop rudimentary shared knowledge structures about the team.
In the context of social coding, and in line with traditional team performance studies, some research suggests that team formation in GitHub is related to previous collaborations and social ties \cite{Casalnuovo15:onboarding}. 

As described in research on teamwork in science and technology, collaboration draws from both social or team factors, as well as technical or task related factors to meet objectives.
More specifically, such work requires an effective integration of both teamwork and taskwork to achieve team goals~\cite{Fiore15:teamwork,Fiore16:techteam}. 
Less research has examined the importance of the taskwork processes established during team formation and its links to team productivity over time. 
To redress this gap, we analyze features characterizing the team formation phase in GitHub repositories.  We examine differences in developer activity-based features extracted during the team formation phase in groups of high- and low-performing software development teams. 

\begin{figure*}
\includegraphics[width=\textwidth]{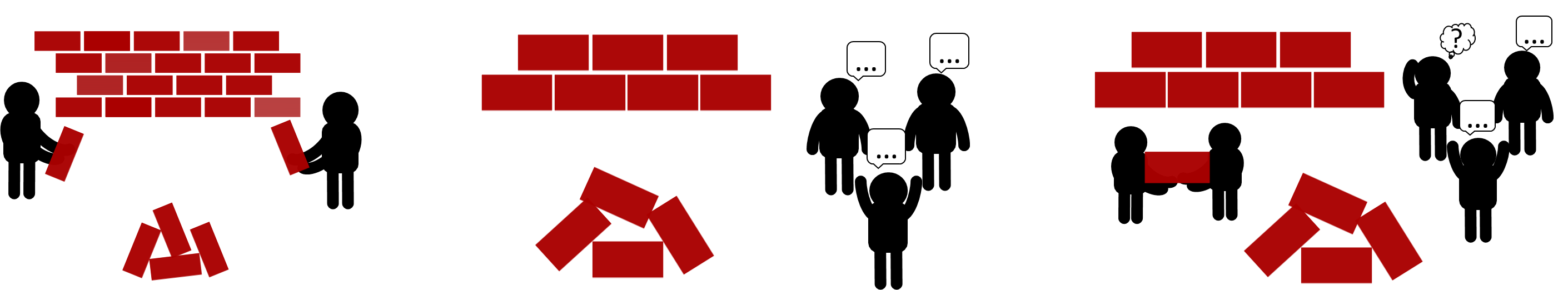}
\caption{From left to right: Toilers, Communicators, and Collaborators}
%\Description{}
\label{fig:teaser}
\end{figure*}

\section{Related Work}

Several threads of research on software development in GitHub are most relevant to the present work. In particular, we survey research on coordination processes, cognitive artifacts, and individual
and group performance evaluation in social coding platforms.
Research on collaborative work in software repositories on GitHub typically examines the activity profiles of developers and/or their responses to surveys and semi-structured interviews to better understand important processes and outcomes of interest.
 This body of work examines differences between contributors, including their varying influence on each other \cite{Blincoe2016:popular}, and their level and length of participation \cite{Joblin:devcoord,vasilescu2015gender,Onoue:devchar}. 
% This section presents a survey of research on 1) coordination processes 2) cognitive artifacts and 3) team performance evaluation in social coding platforms. 
\subsubsection{Coordination Processes}
Coordination can be either implicit or explicit but it generally describes the means through which team members organize their activities towards a shared goal \cite{Rico2008:coord}. These processes shape the team's understanding of interdependencies for individual and collaborative output, leading them to develop strategies for the effective integration of multiple contributions.
In GitHub, implicit coordination takes the form of information-gathering behaviors for the maintenance of task and developer awareness, for example, by reviewing code artifacts or following a developer to receive notifications about their activity \cite{Blincoe2015ruby}.
Explicit coordination is observed in discussions in comments linked to commits, pull requests, and issues.
In a study of highly-watched repositories, research finds that, as team size increases, explicit coordination also increases \cite{Romero:decentralized}.
Other work has similarly found that team size influences coordination structure such that small teams can function without the aid of a designated coordinator while larger teams require top-down management for the coordination of their contributions \cite{Joblin:devcoord}.
Indeed, centralization and delegation of work in software projects is critical for success \cite{Brooks1975:SE} and has been linked to improved issue support quality in software repositories on GitHub \cite{Jarczyk:surgical}.
 
\subsubsection{Cognitive Artifacts}

In the cognitive and social sciences, cognitive artifacts are argued to be integral to effective cognitive and collaborative processes.
Cognitive artifacts are "artificial device[s] designed to maintain, display, or operate upon information in order to serve a representational function" \cite{Norman1991:artifact} and can provide valuable insights about processes and outcomes in distributed cognitive systems \cite{Hutchins1995:cockpit}.
Fiore and colleagues argue that these forms of externalized cognition are central to team cognition, particularly when collaboration occurs through a hybrid human-machine team \cite{Fiore16:techteam}. 
In the present context, software repositories contain a multitude of artifact types, and whether they be code- (i.e., work output) or communication-based (processes), they are devised to support elements of team cognition.
Issue trackers in social coding platforms are official documentation artifacts.
Issue comments, in particular, are artifacts of collaborative problem-solving processes.
Furthermore, through the use of features like issue labels, developers can organize these artifacts.
These features, when consistently used, may support information-gathering activities of current and potential contributors as developers report that they prefer to evaluate task-relevant information in favor of disrupting another working developer \cite{Blincoe2015ruby}. 

\subsubsection{Team Performance Evaluation in Social Coding Platforms}
Performance measurement of software development teams in GitHub is challenging as there are a number of indicators that are related to performance outcomes.
For example, developers report that they measure the success of a software project in terms of the number of contributors and contributor growth \cite{McDonald:performance}. 
Accordingly, studies of software development on GitHub often emphasize the importance of a project's popularity and growth as indicators of its contributors' performance. 
Indicators of repository popularity and growth include its number of stars and forks (copies), respectively. 
Other research identifies  outcomes that can serve as indicators of team performance, including integrator productivity and code quality \cite{Vasilescu:quality}, and issue support quality \cite{Jarczyk:surgical}.
We focus on overall work activity as the primary indicator of team performance.
In this, we examine several different aspects of collaborative work in GitHub: work centralization, work style, and issue support quality.

\section{Method}
Based the prior work described in the previous section, we apply the following criteria to the data collection procedure.
The \textit{team formation phase} was defined to be the first six months following the creation of the repository. We observed the activities of the repositories for a 13 month period after the repositories' creation to assess team productivity; this time period is termed the \textit{evaluation period}.

A user is identified as a \textit{team member} based on their participation in the repository during a specific time period (formation phase or evaluation period). A user is considered a member of the team if they have completed at least one of the following during the team formation phase: one push event; five accepted pull requests; ten issue comments; or ten pull request review comments.

All of the following are treated as work events: (1) push; (2) merged  pull request; (3) issue comment; and (4) pull request review comment.
Furthermore, we consider push events as work output, issue comments and pull request review comments as explicit coordination, and merged pull requests as collaborative work. 

We examined performance or productivity of teams by aggregating work of the evaluation period team at the repository-level.
To view this as a measure of team effectiveness, we averaged productivity over the number of members within a repository.  In other words, we assessed the average amount of work completed by individual team members 13 months after the creation of the repository.

\subsection{Data Set}
The data set used in this study contains all GitHub events from January 2016 to June 2017.
From this data set, we selected software repositories created in January 2016 and included only those that had more than 20 work events and at least two members in the team formation phase. 
% Our data set included 20,950 active repositories and the 96,447 unique GitHub users contributing to those repositories in the team formation phase.
We measured size of the teams, once more, in the evaluation period and removed the repositories with less than two members. 
Our final data set included 20,370 active repositories and the 59,178 unique GitHub users contributing to those repositories in the evaluation period.

\subsection{Team Feature Extraction}

We extracted team features based on event data in the team formation phase.
\begin{itemize}
    \item \textbf{Relative frequency of work events} 
    is the ratio of the number of each work event to the total number of work events. This feature allows us to evaluate the type(s) of work undertaken during team formation. 
    \item \textbf{Work events per person} 
    is the number of each work event divided by the number of team members. We used this feature to evaluate the amount of work, or productivity of contributors.
    \item \textbf{Burstiness} 
    measures the temporal correlation of activities within a team and defined as equation \ref{eq:burstiness}; where $\mu_{\tau}$ and $\sigma_{\tau}$ are mean and standard deviation wait times $P(\tau)$.
    \begin{equation}
       Burstiness=\frac{\sigma_{\tau} - \mu_{\tau}}{\sigma_{\tau} + \mu_{\tau}} \label{eq:burstiness}
    \end{equation}
    An analysis of burstiness reveals the presence of increased, synchronized activity in a team and is linked to effective collaborations and improved team performance \cite{riedl2017teams}.
    \item \textbf{Issue labeled proportion} 
    is the proportion of issues in the repository that are labeled. Because issue labels are one way that developers can organize their work and communication on GitHub, we use this feature to evaluate the use and organization of cognitive artifacts by the team.
    \item \textbf{Team size} 
    is the number of team members in the repository during the team formation phase. This feature was used to assess the importance of the team's size early on in its development.
\end{itemize}

% \subsection{Performance Prediction}

% We defined the performance of teams based on their productivity a year after the creation of their software project repository. Due to the sparsity of events in our repository data, a month was selected as the time period used to evaluate performance. Because popularity as measured by number of stars is commonly used as a measure of success in software projects on GitHub, we calculated the Pearson correlation coefficient for the number of stars and log work events per person, and found a high correlation (\textit{r} = 0.77). We thus deemed our productivity measure to be sufficiently linked to project performance and representative of effective team collaboration.

% We create a predictive model for team performance

% To build a model to predict the performance of teams during the evaluation period based on their features extracted in the team formation phase, we used the Random Forest classifier in the \textit{scikit-learn} library \cite{scikit-learn}.
% We trained the model using 200 estimators and set the maximum depth of the tree to eight to avoid overfitting.

\section{Results}

\subsection{Team Performance Evaluation}

First, we examine the characteristics of the work event data during the performance evaluation period. Performance is defined as the amount of the work teams completed per person in the evaluation period.  Figure \ref{fig:WorkHist}A shows that the distribution of work per person is a heavy-tailed distribution. A log transformation on work per person was applied to decrease the variability of this measurement. The transformed distribution is a normal distribution with a mean and variance of 2.05 and 0.37, respectively (Figure \ref{fig:WorkHist}B).
We considered the log transformed values of work per person representative of team productivity. To categorize teams based on their productivity, we used the median of team productivity (= 2.0) as our threshold and labeled teams as high-performing if their productivity was greater than the threshold and low-performing otherwise.

\begin{figure*}
    \centering
    \includegraphics[width=0.8\textwidth]{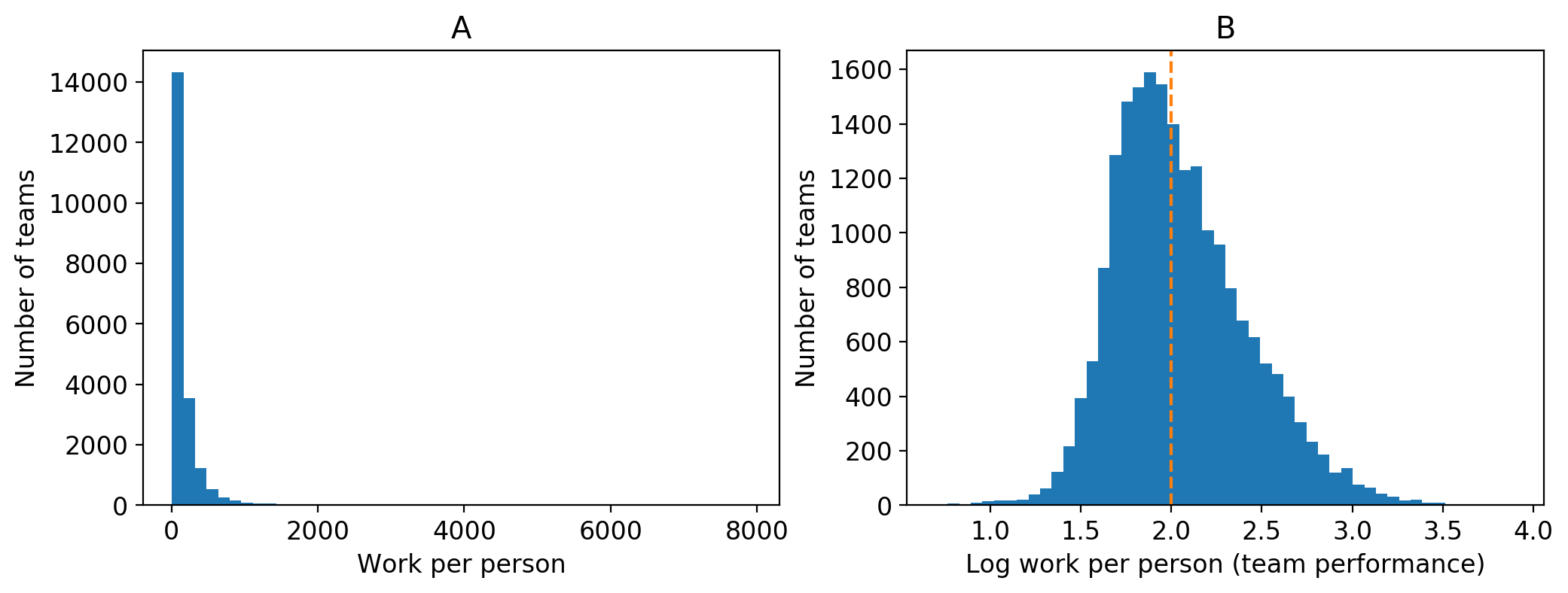}
    \caption{Distribution of work per person per month (A) and transformed distribution of work per person per month (B).}
    \label{fig:WorkHist}
\end{figure*}

% \subsubsection{Evaluation Period Teams}

Our unit of analysis is based, in part, on the size of the team during the performance evaluation period. 
% We found that 562, approximately 3\%, of the  repositories that were active in the team formation phase were no longer active in the performance evaluation period. These are likely short-term repositories, such as school projects. Moreover, 18 repositories lost the majority of their members and their size decreased to one. We removed these repositories from our analysis as we are focused on studying collaborative work in long-term software projects.
Figure \ref{fig:TeamSizeHist} provides the histogram of team sizes in our data during the performance evaluation period.

\begin{figure}[h]
  \centering
  \includegraphics[width=0.8\linewidth]{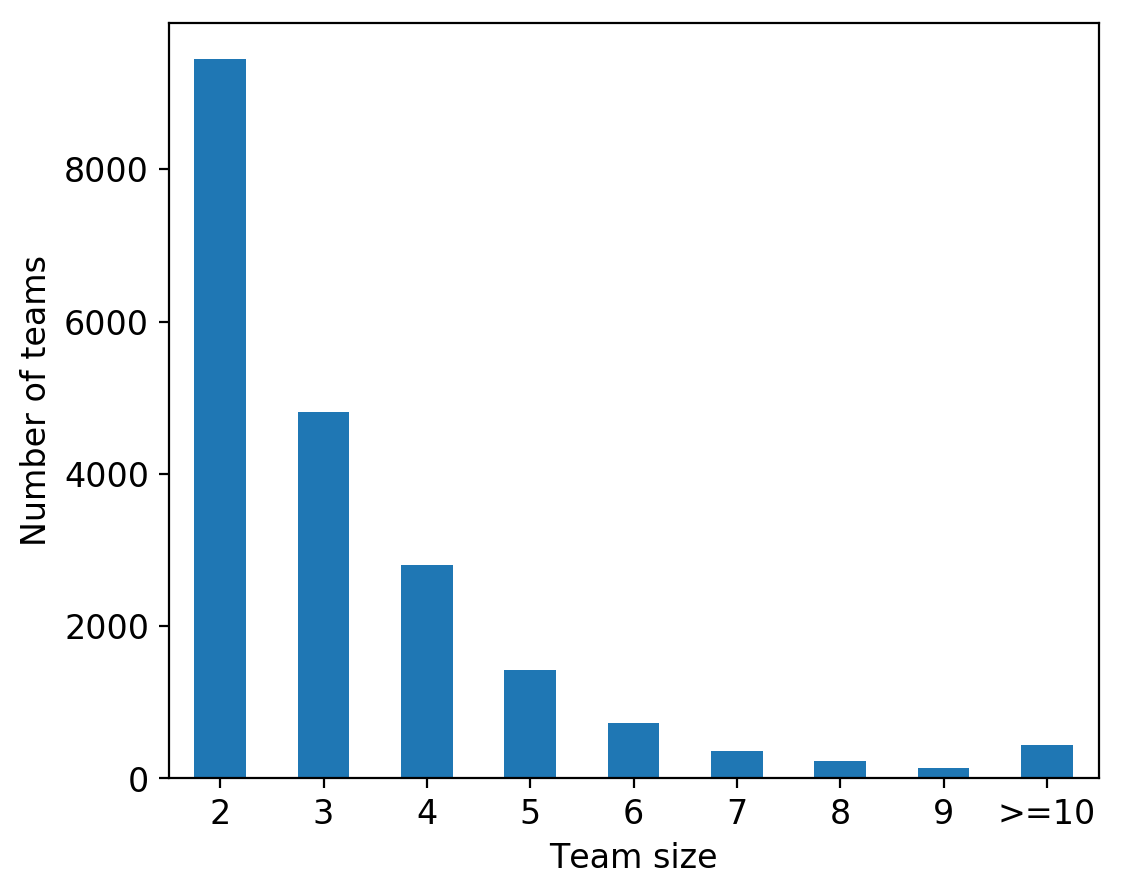}
  \caption{Team sizes in the performance evaluation period.}
  \label{fig:TeamSizeHist}
\end{figure}

\subsubsection{Evaluation Period Team Sizes}

For additional analyses, we grouped teams based on their size: teams with three or fewer members were considered small, teams with between four and six members were considered medium, and teams with seven or more members were categorized large. Vasilescu et al. \cite{vasilescu2015gender} used higher thresholds to categorize teams in software repositories on GitHub. However, given the short life span of the repositories we studied, we determined these thresholds were more appropriate to capture differences in processes and outcomes across team sizes.  The number and the proportion of teams in each group is provided in Table \ref{tab:TeamSize}. In the performance evaluation period, more than half of the repositories in our data set were maintained by small teams and less than a quarter were maintained by medium-sized teams. Large-team repositories made up less than five percent of the data.

\begin{table}[]
    \centering
    \begin{tabular}{|c c c c|}
          \hline
         Group & Size & Proportion & Frequency\\ 
         \hline\hline
         Small & [2, 3] & 70\% & 14261\\ 
         \hline
         Medium & [4, 6] & 24\% & 4951 \\
         \hline
         Large & [7, inf) & 6\% & 1157 \\
         \hline
    \end{tabular}
    \caption{Proportion of teams by team size.}
    \label{tab:TeamSize}
\end{table}

Figure \ref{fig:TeamSizePerformance}A shows the performance of teams of different sizes; teams with two members have slightly higher performance compared to teams with three, four, or five members. 
Given that most contributors do very little work in GitHub repositories \cite{Vasilescu2014:workload}, this finding is not particularly surprising. 
Figure \ref{fig:TeamSizePerformance}B shows the number of high- and low-performing teams in each team size category. 
Interestingly, there are more high performers than low performers in the large team size group. This suggests that, in our data set, large teams are likely engaged in more frequent interactions and potentially more effective collaborations leading to the production of more communications and code artifacts (i.e., more work events per person).

\begin{figure*}
    \centering
    \includegraphics[width=0.8\textwidth]{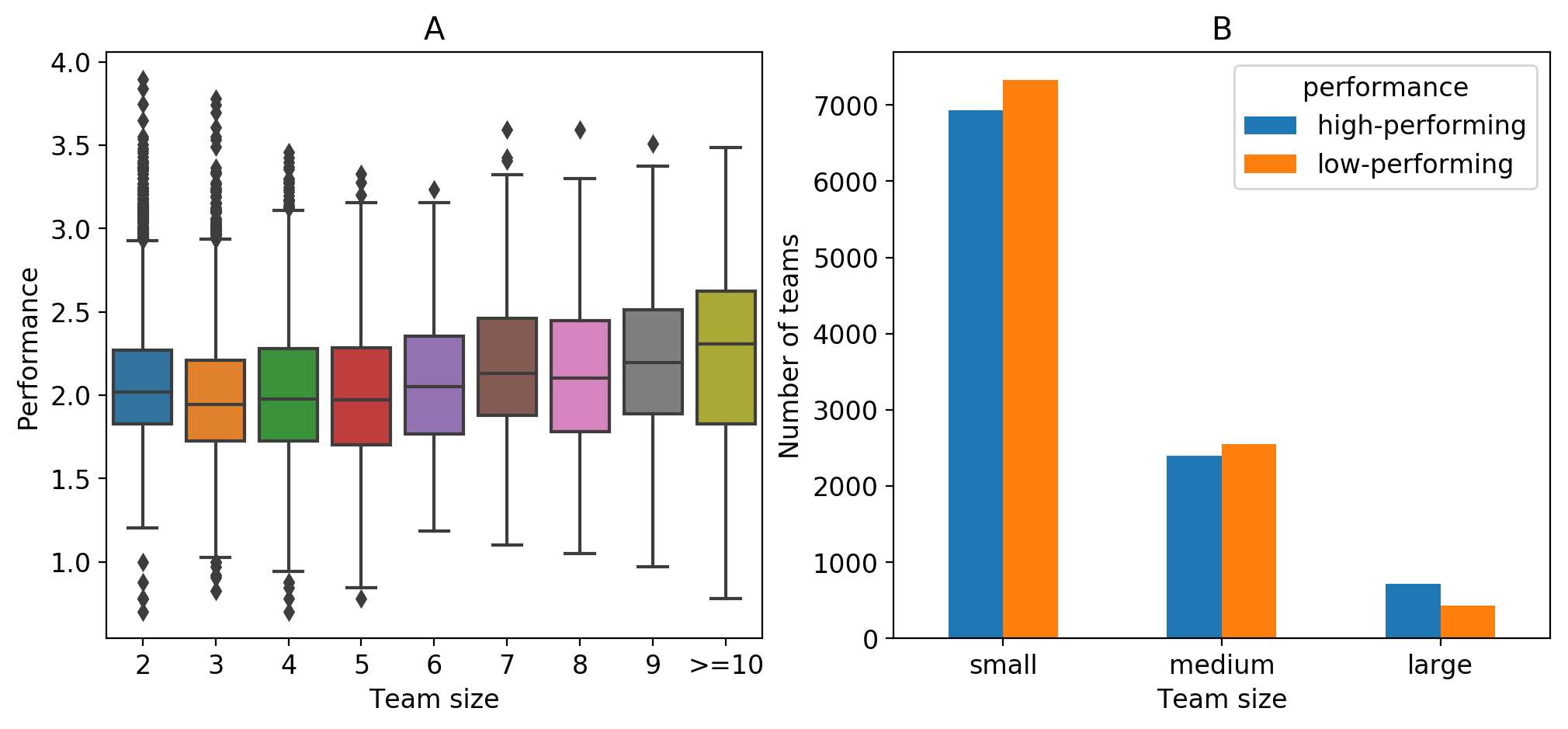}
    \caption{ Performance of teams by size (A) and high- and low-performing teams by team size groups (B).}
    \label{fig:TeamSizePerformance}
\end{figure*}

\subsection{Work Centralization}

To examine the distribution of work underlying the performance differences observed across team sizes, we calculated the amount of work centralization in teams using the Gini coefficient.
The Gini coefficient is a measure of inequality, where 0 represents perfect equality and 1 represents maximum inequality.
We use it to analyze the inequality of work events per person.
More specifically, a higher Gini coefficient suggests that a smaller set of team members do most of the work; that is, work is centralized to a fewer number of members in the repository.
Figure \ref{fig:Gini} provides distributions for the Gini coefficient values of teams in the performance evaluation period.
High-performing teams have a higher Gini coefficient compared to low-performing teams, regardless of team size.
The difference between the Gini coefficient of low- and high-performing teams is greater in large teams compared to small and medium teams.
This finding provides support for the importance of work centralization for performance in OSSD in GitHub \cite{Jarczyk:surgical}.
Our results suggest that work centralization is linked to increased overall productivity in software repositories and is particularly important for collaborations in large teams.

\begin{figure*}
    \centering
    \includegraphics[width=0.8\textwidth]{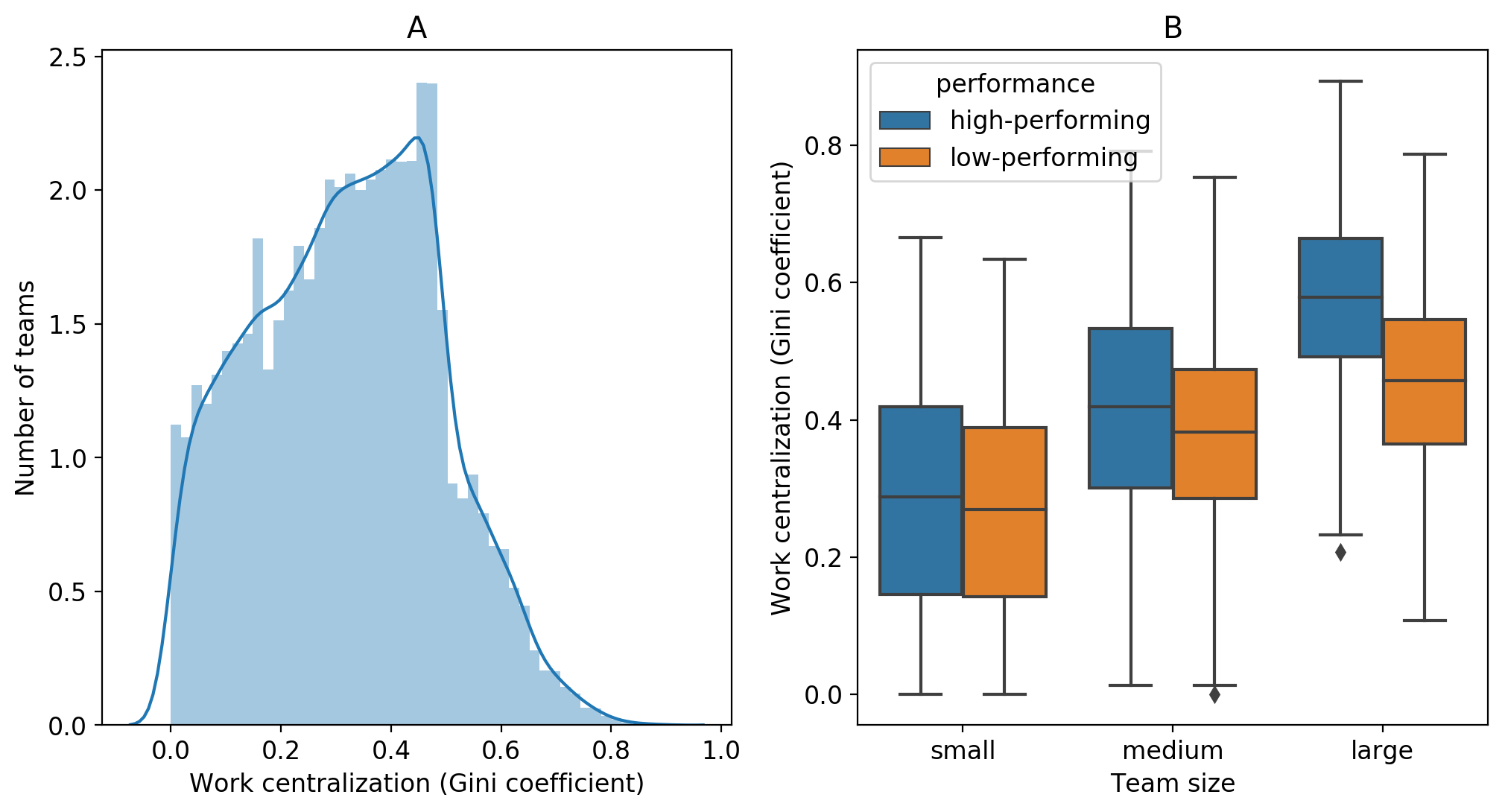}
    \caption{Distribution of Gini coefficients for all teams (A) and distribution of Gini coefficients by team size and performance (B).}
    \label{fig:Gini}
\end{figure*}

\subsection{Work Style Clusters}
GitHub users and teams behave differently and clustering them can express their diversity \cite{saadat2018initializing}. To discover various work styles on GitHub, we clustered teams using the \textit{k-means} clustering algorithm applied to the proportion of different types of work events. 
We tested various $k$ values for the \textit{k-means} clustering and observed that $k=3$ generates the minimum number of clusters that are meaningfully distinct from each other.
The number of teams in each of the three work style clusters is provided in Table \ref{tab:WorkStyleClusters}, and the cluster centroids are plotted in Figure \ref{fig:Clusters}.
Based on the proportion of events in work style clusters, we labeled them as: \textit{toilers}, \textit{communicators}, and \textit{collaborators}. 

\begin{figure}
    \centering
    \includegraphics[width=\linewidth]{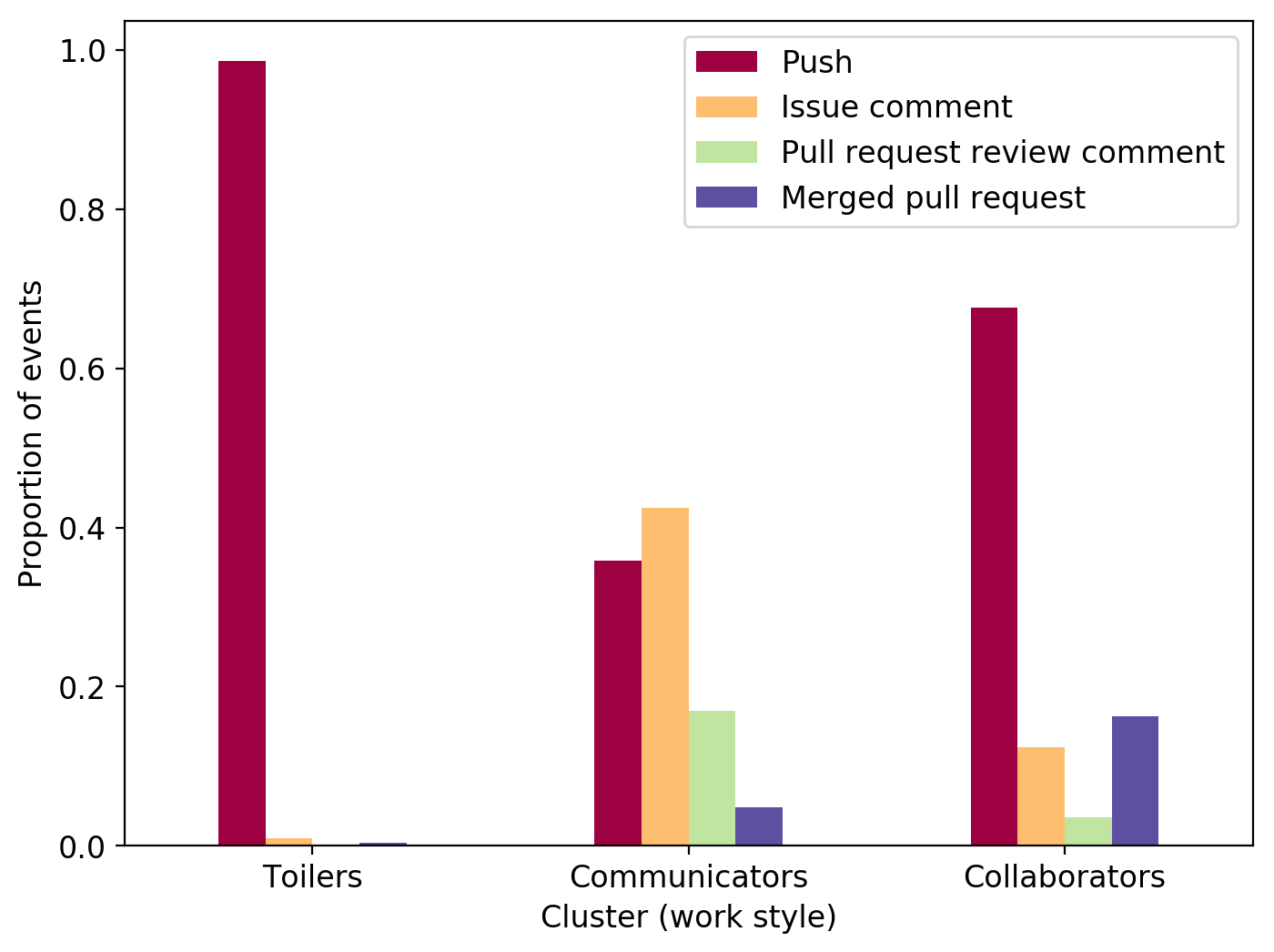}
    \caption{Proportions of events by work style cluster.}
    \label{fig:Clusters}
\end{figure}

\textit{Toilers} produce a higher proportion of push events compared to \textit{communicators} and \textit{collaborators} but have limited, if any, communication taking place in GitHub. 
This may reflect a failure to engage in explicit coordination but it is also possible that these teams use alternative communication channels, like Slack \footnote{https://slack.com/} and Discord \footnote{https://discordapp.com/}, to coordinate their contributions, as past research has shown \cite{Storey17:channels}. 
Although \textit{toilers} focus on code contributions, they generally do not accept external contributions, unlike the other two work style groups. \textit{Communicators} produce a higher proportion of comment events, and \textit{collaborators} produce a higher proportion of pushes and merged pull requests.

For additional analysis, we calculated the average number of each event type for different work styles (Table \ref{tab:WorkStyleClusters}).
Overall, \textit{collaborators} have the highest level of productivity in terms of internal and external code contributions (pushes and merged pull requests, respectively) whereas \textit{communicators} have the highest level of productivity in terms of explicit coordination.
Although \textit{toilers} devote the majority of their effort to code contributions, they still have, on average, a lower number of pushes compared to \textit{collaborators}. 

\begin{table*}[]
    \centering
    \begin{tabular}{|l l | l | l | l| }
    \hline
    Work style cluster &  & \textit{Toilers} & \textit{Communicators} & \textit{Collaborators} \\
    \hline
        Number of teams& & 15264 & 3219 & 1886 \\
        \hline
        \multirow{2}{*}{Push} & mean & 79.61 & 68.61 & 91.60 \\
         & std & 118.19 & 143.40 & 128.98 \\
        \hline
        \multirow{2}{*}{Merged PR} & mean & 0.55 & 14.93 & 23.12 \\
        
         & std & 4.53 & 47.79 & 45.15 \\
        \hline
        \multirow{2}{*}{Issue comment} & mean & 0.86 & 109.72 & 17.31 \\
        
         & std & 4.61 & 477.73 & 35.62 \\
        \hline
        \multirow{2}{*}{PR review comment} & mean & 0.09 & 41.12 & 5.65 \\
        
         & std & 1.51 & 142.60 & 19.54 \\

        \hline
    \end{tabular}
    \caption{The mean and standard deviation of different work events for toilers, communicators, and collaborators.}
    \label{tab:WorkStyleClusters}
\end{table*}

\subsubsection{Work Style and Performance}

Figure \ref{fig:ClustersPerformance} shows the number of high- and low- performing teams in each work style cluster.
Nearly 75\% of the teams in our data are in the \textit{toilers} cluster. Of this subset, the majority of them were low-performing teams.
This suggests that a lack of communication within GitHub is indicative of poor performance.
Although \textit{toilers} could be communicating outside the GitHub ecosystem, the fact that they did not perform relatively well suggests that direct communication within GitHub (e.g., issue comments) may help overall team performance.
This is illustrated when we look at the \textit{communicators} and \textit{collaborators} clusters.
For the \textit{communicators}, we see a relatively more equal distribution of comments proportional to the amount of work done.
For the \textit{collaborators}, there is relatively more communication compared to the toilers.
And, proportionate to the amount of other activity, they also accept more pull requests than any of the other clusters.
In the \textit{communicators} and \textit{collaborators} clusters, the majority of teams are high performers.
This effect is more pronounced for large teams and, to a lesser extent, medium teams compared to small teams. 

\begin{figure*}
    \centering
    \includegraphics[width=\linewidth]{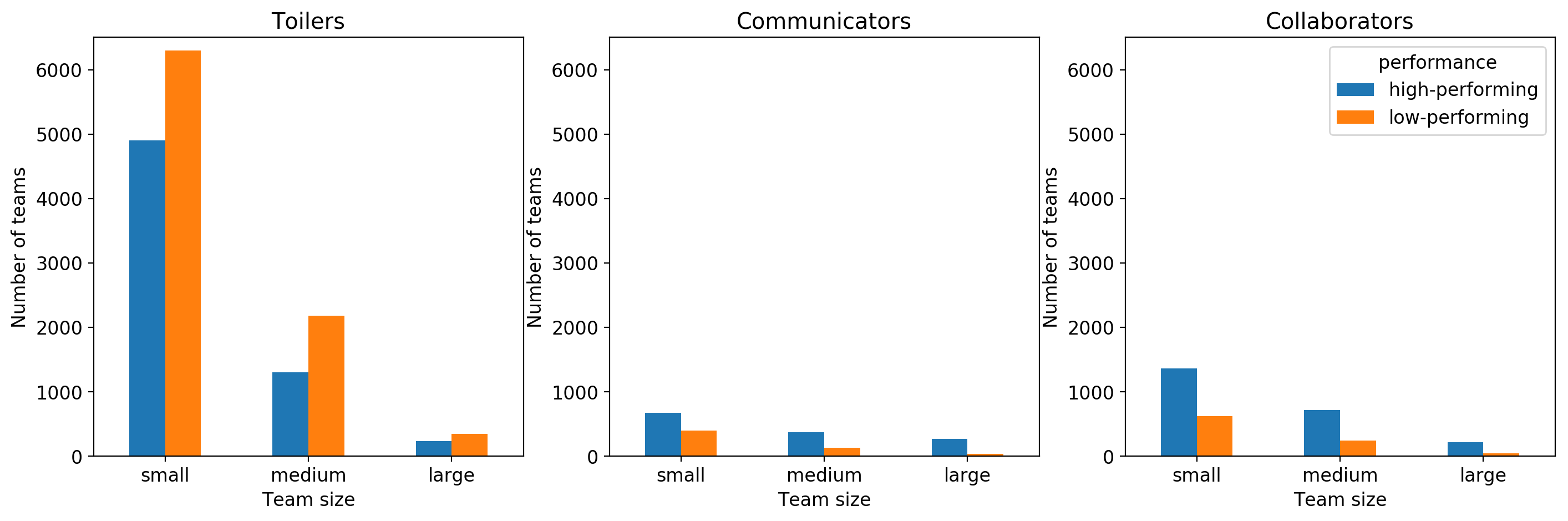}
    \caption{The performance of different work styles and team size groups.}
    \label{fig:ClustersPerformance}
\end{figure*}

\subsection{Team Features}

Our goal is to understand the impact of team formation phase on the evaluation period.
Figure \ref{fig:Features} illustrate the team formation phase features that are linked to performance of teams in the evaluation period; that is, the features predictive of success. 
Across each of these, t-tests show that the \textit{p-value} is less than $10^{-10}$ in all features represented in figure \ref{fig:Features}.
More specifically, these features identified during the formation phase, and indicative of the type of work norms created during formation, are significantly related to higher performance during the evaluation phase.

\subsubsection{Types of Activity}

Large-team repositories have a higher proportion of coordination events (i.e., issue comments in figure \ref{fig:Features}B) and collaborative work events (merged pull requests in figure \ref{fig:Features}C). Conversely, these repositories have a lower proportion of internal contribution events (pushes in figure \ref{fig:Features}A). 
High-performing teams have a lower proportion of push events and a higher percentage of coordination events.
In other words, high-performing teams exhibit higher levels of coordination.
Issues in GitHub repositories can be used to develop an understanding of the problem at hand and the work needed to resolve the problem, and also offer a space for the discussion and evaluation of potential solutions and strategies to address needs of users and the project.
The larger proportion of issue comments in high-performing teams shows that dissections of issues are helpful in improving the productivity of the teams. 
In contrast, low-performing teams have a lower proportion of coordination events and a higher proportion of contribution events.
This suggests that these teams are primarily focused on their work output and spend less time on coordination and communication.
Fitting with the team cognition literature \cite{Fiore10a:macrocog}, this may lead to a failure to evaluate alternatives and the premature selection of solutions.

\subsubsection{Burstiness}

Overall, we observe a higher amount of burstiness in high-performing teams and a lower amount of burstiness in low-performing teams.
Interestingly, this difference in high- and low-performing teams is consistent across team sizes.
This suggests that high-performing teams, in general, interact more frequently and their activities are highly-synchronized. 

\subsubsection{Issue Labels}

Figure \ref{fig:Features}F shows that proportion of labeled issues is generally higher for high-performing teams than it is for low-performing teams. 
This is particularly true for medium and large teams and less noticeable in small teams.
Again, this finding supports the claim that larger teams require more coordination mechanisms to function effectively.
This suggests that, as a classification system for artifacts, the consistent use of issue labels may scaffold collaborative problem-solving processes \cite{Fiore16:techteam} and thus support the productivity of software development teams in GitHub.
More specifically, in line with team cognition theory on complex problem solving \cite{Fiore10b:macrocog}, issue labels provide support for information-gathering and knowledge-building activities of current and prospective team members. 

\begin{figure*}
    \centering
    \includegraphics[width=\textwidth]{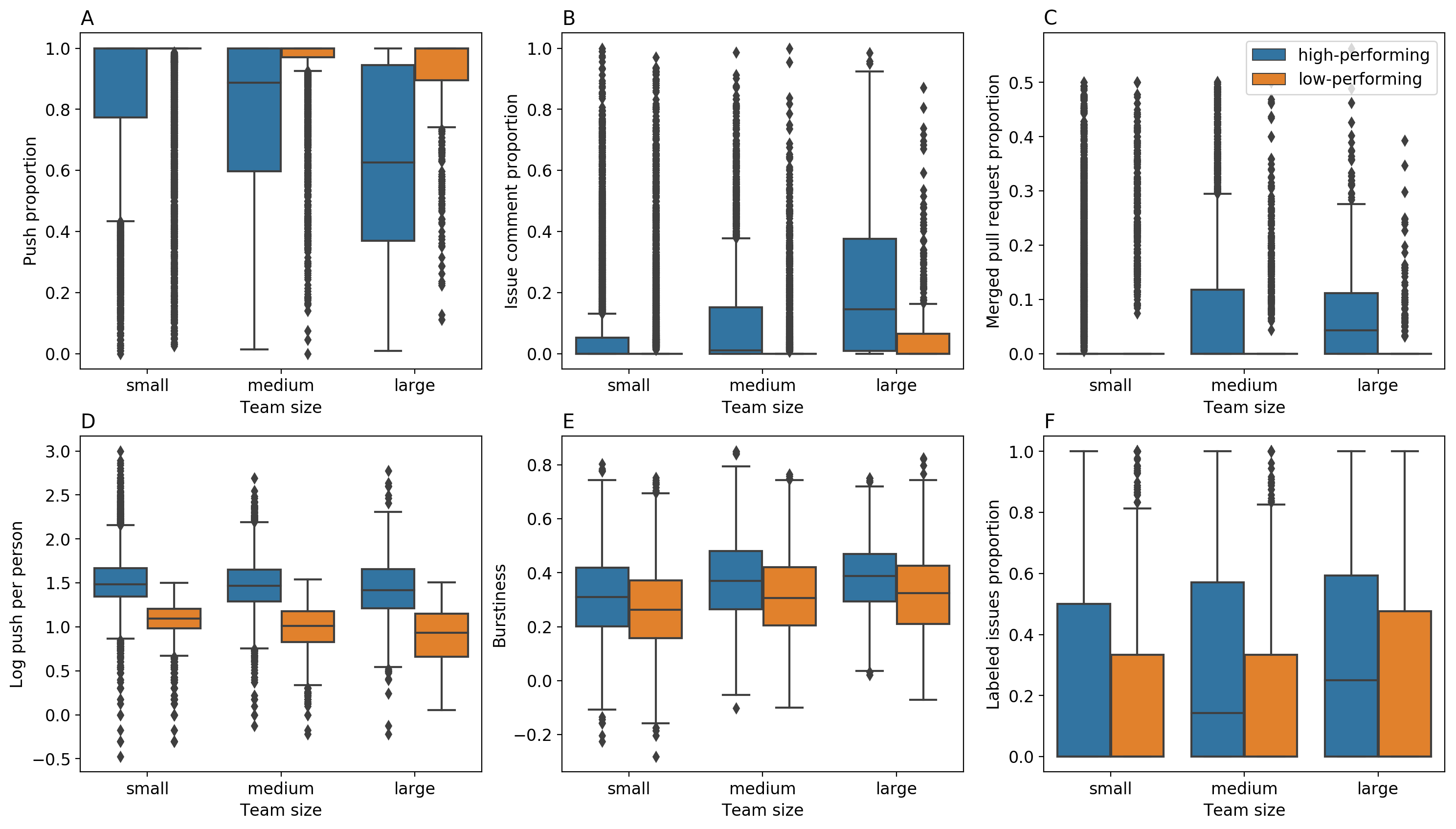}
    \caption{Team formation phase features by team size and performance group.}
    \label{fig:Features}
\end{figure*}

\section{Limitations and Future Work}

Given the complicated nature of OSSD, software repositories have different development models.  As described, in the fork \& pull model, developers fork the main repository, push their changes to the forked repository, and when their task is completed, the developer sends a pull request to the maintainers of the main repository and ask them to accept their changes. In this study, we did not consider the additional push events performed on the forked repositories. This could lead to a lack of accounting for a portion of push events in fork \& pull-based repositories. 

Another issue is how bots influence team collaborative activities. Bots are increasingly common in OSSD \cite{Lebeuf18:bots}. We did not identify and remove bots from our data set. It is possible that some differences in productivity may be influenced, in part, by bot activity. Some developers also use multiple accounts on GitHub; we did not apply any aliasing to usernames to identify multiple accounts used by a single user. 

Finally, we did not consider specifics of work centralization across users.
In future work, we plan to examine performance differences in core and periphery contributors as it relates to work centralization and overall productivity. The core and periphery are qualitatively distinct in GitHub, contributing at different rates and exhibiting different levels of interdependency \cite{Joblin:devcoord,Joblin2017:core}. Additionally, we plan to develop and evaluate a combinatory metric for team performance in GitHub based upon factors such as work centralization and other factors indicative of efficiency and effectiveness. This will allow us to holistically examine the multidimensionality of team performance in social coding platforms.

\section{Conclusion}

% We built a model to predict GitHub teams' performance based on their team formation phase activities.
We studied the impact of team formation phase activities on the performance of the team in the future.
Our results show that coordination processes manifest themselves as particular GitHub features, and that the diversity of work in GitHub repositories can be linked to productivity.
This work makes both theoretical and practical contributions.
At the theoretical level, our findings contribute to the body of work in team cognition by showing how features relating to teamwork and taskwork, and their occurrence during team formation, influence performance at later development times.
It also documents the importance of coordinating mechanisms, like cognitive artifacts in GitHub, and how they are related to team effectiveness.
At the practical level, understanding the features we extracted, and how they relate to team performance, could help repository maintainers and developers lead their efforts more efficiently.
When creating new projects, developers who would like to attract productive contributors could consider how their use of the platform for coordination purposes will influence the type and levels of work that will emerge over time.
When identifying new projects to contribute to, developers may consider the amount of explicit coordination they observe in the repository to gauge the work style and productivity levels of the developers in the repository.

\section{Acknowledgements}
This work was supported by grants FA8650-18-C-7823 and W911NF-20-1-0008 from the Defense Advanced Research Projects Agency (DARPA). The views and opinions contained in this article are the authors and should not be construed as official or as reflecting the views of the University of Central Florida, DARPA, or the U.S. Department of Defense.

%
% The next two lines define the bibliography style to be used, and the bibliography file.
\bibliographystyle{IEEEtran}
\bibliography{main}

\end{document}